# The onset of planet formation in brown dwarf disks

**Dániel Apai[1,2,3,*], Ilaria Pascucci[1,3], Jeroen Bouwman[4], Antonella Natta[5],**

**Thomas Henning[4], Cornelis P. Dullemond[4]**

[1]*Steward Observatory, The University of Arizona, 933 N. Cherry Avenue, Tucson, AZ-85721, USA.* [2]*NASA Astrobiology Insitute.* [3]*These authors contributed equally to this work.* [4]*Max Planck Institute for Astronomy, Königstuhl 17, Heidelberg, D-69117, Germany.* [5]*Osservatorio Astrofisico di Arcetri, INAF, Largo E. Fermi 5, I-50125 Firenze, Italy*

* *To whom correspondence should be addressed. E-mail: apai@as.arizona.edu*

**The onset of planet formation in protoplanetary disks is marked by the growth and crystallization of sub-micron-sized dust grains accompanied by dust settling toward the disk mid-plane. Here we present infrared spectra of disks around brown dwarfs and brown dwarf candidates. We show that all three processes occur in such cool disks in a way similar or identical to that in disks around low- and intermediate-mass stars. These results indicate that the onset of planet formation extends to disks around brown dwarfs, suggesting that planet formation is a robust process occurring in most young circumstellar disks.**

Planet formation starts with the growth of sub-micron sized amorphous grains in the protoplanetary disks (e.g. *1,2*). Theory predicts that larger grains settle faster to the disk mid-plane resulting in flattened disk geometries (e.g. *2,3*). Observational evidence for dust settling has been found for disks around young low-mass (T Tauri) and intermediate-mass (Herbig Ae/Be) stars (e.g. *4*). Significant dust processing in the early Solar System is demonstrated by the high crystallinity of some comets containing dust from the epoch of their formation (*5*). Recently, intermediate-mass stars were shown to have high crystallinity only when larger grains are present (*6*), suggesting a possible link between grain growth and crystallization. Up to now, detailed dust composition studies were limited to bright disks of intermediate-mass stars which suggested very low or no crystallinity for disks around low-mass stars (e.g. *6*). If true, dust processing would strongly depend on the stellar properties and planet formation processes would differ or not occur at all in disks of very low–mass stars. Recently, there is accumulating evidence that crystalline silicates are present in disks of low-mass stars (e.g. *7,8*). Even more surprisingly, ground-based photometry of a brown dwarf disk showed hints for grain growth and dust settling (*9*), and crystalline silicate features were identified in the disk of a brown dwarf candidate (*10*). These findings set the question whether such few Jupiter mass-disks (*11,12*) can form planets. In this report, we present mid-infrared spectra of disks around very low-mass young stellar and substellar objects. We show that five out of six disks have highly processed dust: large grains and very high crystalline mass fractions (~40%) are found. The correlation between the shape and strength of the silicate emission feature observed for Herbig Ae/Be disks extends to brown dwarf disks, demonstrating that dust processing is independent of the stellar properties. All the disks with highly processed dust have strongly flattened disk structure, as expected from dust settling. We conclude that the first steps of planet formation occurred in these brown dwarf disks, suggesting that even substellar disks can form planets.

We used the Spitzer Space Telescope and its sensitive Infrared Spectrograph to survey the complete population of substellar-mass objects with previously identified mid-infrared excess emission (*13,14*) in the Chamaeleon I star-forming region. We obtained low-resolution ($\lambda/\Delta\lambda{\sim}60$-120) infrared spectra between 7.7 and 14.4 μm, covering the 10 μm silicate emission feature, whose shape and strength is determined by dust grain size and composition. Our targets have been spectroscopically classified as brown dwarf candidates or objects on the stellar/substellar boundary (*13, 14, 15,16*). The observations provide a yet unique, coeval sample (1–3 Myr) of cool objects with temperatures between 2500 – 3100 K, expanding the range in stellar mass over which dust composition has been studied to two orders of magnitude, a factor of two in temperature and four orders of magnitude in luminosity. The spectra were taken using multiple ramp cycles and reduced with the SMART reduction package and routines we developed (*16,17*). We confirm mid-infrared excess emission — indicative of disks — for six of our targets; the fluxes of the two other objects (Cha 449 and Cha 425) are consistent with pure photospheric emission and are excluded from further analysis.

The spectra allow morphological comparison with the infrared spectrum of the interstellar medium and that of comet Hale-Bopp, as shown in Fig.1. All six brown dwarf disks have emission features substantially broader than that of the interstellar medium, indicative of larger grains (e.g. *18*). While the spectrum of Cha Hα1 peaks at 9.8 μm, similarly to the dominantly amorphous interstellar grains, the other five targets show prominent crystalline silicate emission features with characteristic peaks at 9.3 μm and 11.3 μm. In particular, the spectrum of Cha 410 resembles that of comet Hale-Bopp. The faintest of our targets, Cha

Hα9, shows a strong crystalline contribution, but due to its low signal-to-noise ratio we excluded it from the further quantitative analysis. To link grain growth and crystallization (7,19), in Fig. 2 we study the relation between the strength and the shape of the silicate feature by plotting the flux ratio at 11.3 μm and 9.8 μm against the peak over continuum flux ratio. To ensure homogeneity, we used a single procedure to derive these values for young intermediate-mass stars (6), low-mass stars (7) and for our brown dwarf sample. For intermediate-mass stars a linear correlation has been proposed by (6): weaker features have more crystalline contribution. We show that this correlation is also valid to disks around low-mass stars as well as for brown dwarf disks. Based on simulating a silicate emission feature and altering its dust composition, we also plot three vectors indicating small grain removal, grain growth and grain crystallization. The location of our targets in the plot demonstrates that dust processing (grain growth, grain removal and/or crystallization) occurred in all of them. The fact that the disks around intermediate-, low-, and substellar mass objects follow the same correlation shows that the dust processing is very similar or identical in these systems. Due to the nature of the plotted quantities the correlation becomes non-linear for highly processed dust (flux at 11.3 μm ≥ flux 9.8 μm), when the crystalline emission feature becomes dominant. We suggest that for very strongly processed grains the feature strength may increase with increasing crystallinity and will reverse the observed correlation. The plot demonstrates that high crystallinity is always accompanied by significant grain growth for stars of all masses, suggesting a direct link between these processes and very similar time scales. The reliable age estimate of our clustered targets proves that significant dust processing occurs as rapidly as 1–3 Myr. Such rapid dust processing argues for either efficient radial mixing, if the dust is crystallized in the inner disk (20); or, for the early heating of the disk through accretion leading to crystallization at larger disk radii.

We quantify the dust processing  by decomposing the observed spectra into emission from five dust species (*16*, Fig. S2), following the method applied for intermediate-mass stars (*6,18*). For comparison purposes we opted to use the same five major dust species (amorphous: olivine, pyroxene, silica; crystalline: forsterite, enstatite), each with two grain sizes (0.1 µm and 1.6 µm radius) to allow characterization of crystallinity and grain growth in the upper disk layer observable in the mid-infrared. The quantitative study of the five brown dwarf disks confirms the morphological comparison of the spectra: four of the sources have significant contribution from crystalline silicates (9 – 48% mass fractions) and from large grains. As a next step we plot the crystalline mass fractions in disks of brown dwarfs, of low- (*21*) and of intermediate-mass (*6*) stars against the (sub)stellar mass and temperature  (see, Fig. 3). Four out of five disks have high crystalline mass fractions compared to the disks of intermediate and low-mass stars. Remarkably, the previously suggested trend of crystallinity increasing with the stellar mass (predicting little or no crystallinity for low-mass stars, *6*) is not valid. Our data shows a slight increase in crystallinity with decreasing stellar temperatures. This trend can be interpreted keeping in mind that infrared observations sample smaller radii for the cooler disks than for the warmer ones. If  crystals form in the inner disk and diffuse outward via turbulence in a similar way in all disks, probing the inner disks will result in higher crystalline contribution. Apart from this slight increase no obvious, direct relation exists between the stellar temperature  and crystallinity. We conclude that the crystallinity is largely independent of  the stellar properties and it is likely determined by local processes in the disk, such as the efficiency of  radial mixing.

We find an apparent anti-correlation of crystalline mass fraction with the age of the objects, also noted by *(6)* for intermediate-mass young stars; however, the large uncertainties in the age estimates for the isolated intermediate-mass stars do not allow yet any firm conclusions *(17)*.

While the silicate emission feature is determined by the dust grain properties, the continuum of the spectrum is defined by the disk geometry. In Fig. 4 we plot the flux densities outside the silicate emission feature (at 8 and 13 μm) and compare them to the emission from flat and flared disk models *(22)*. Disks with "flared" geometry (disk opening angle increasing with the radius) intercept more stellar light leading to steeply increasing emission toward longer wavelengths. Maintaining the "flared" geometry requires turbulent gas to keep the dust from settling to the disk mid–plane. Disk models suggest that with increasing dust grain size the gas – dust coupling becomes ineffective, leading to the gradual settling of the dust toward the mid–plane and a 'flatter' disk structure. In this evolutionary picture, all gas-rich disks start with flared geometry and will evolve to flat disks as the result of dust evolution. In agreement with this sequence, we find that the five disks with highly processed dust have intermediate or flat disk geometries, demonstrating that dust settling has occurred similarly to disks of low-mass stars *(e.g. 4)*. The sixth object, Cha Hα1, is closer to the flared disk model, consistent with the less processed grains as derived in Fig. 2. The identified dust settling implies that dust grains larger than those observed in the upper, optically thin disk layer are present close to the disk mid–plane *(23)*.

Here we showed that grain growth, crystallization and dust settling have occurred in brown dwarf disks. It appears that the first steps of planet formation are very similar or identical in disks around intermediate- and low-mass stars, and even in the substellar regime. We suggest that the central star does not play key role in the planet formation steps beyond grain growth: large grains (> 100 μm) or small planetesimals will evolve independently of the central object. It seems thus likely that planetesimals, and eventually planets will also form in brown dwarf disks, underlining the robustness of planet formation. We speculate on the expected planetary architectures by scaling the minimum-mass solar nebula ($10 - 70$ Jupiter mass, *24*) to the mass of the two brown dwarf disks with measured masses (*12*). The minimum-mass solar nebula gives birth to an ensemble of planets with one Jupiter mass and below: if the brown dwarf disk mass is distributed in a similar fashion, the most massive planets are expected to be Neptune-like. There should be enough material present to form terrestrial planets in the inner disks, and therefore the closest brown dwarfs should be important targets for future planet searches.

25. We are indebted to R. van Boekel and F. Przygodda for providing their data sets on Herbig Ae/Be and T Tauri stars. We thank G. Rieke, M. R. Meyer, M. Silverstone, and P. Apai for discussions. This material is based upon work supported by the NASA through the NASA Astrobiology Institute under Cooperative Agreement No.

CAN-02-OSS-02  issued through the Office of Space Science. This work is based on observations made with the Spitzer Space Telescope, operated by the Jet Propulsion Laboratory, California Institute of Technology under NASA contract 1407. Support for this work was provided by NASA through Contract Number 1268028 issued by JPL/Caltech. The MPIA team acknowledges support from the European Community's Human Potential Programme under the contract HPRN-CT-2002-00308,PLANETS. We acknowledge the constructive and helpful comments of the two referees which substantially improved the clarity and presentation of this work.

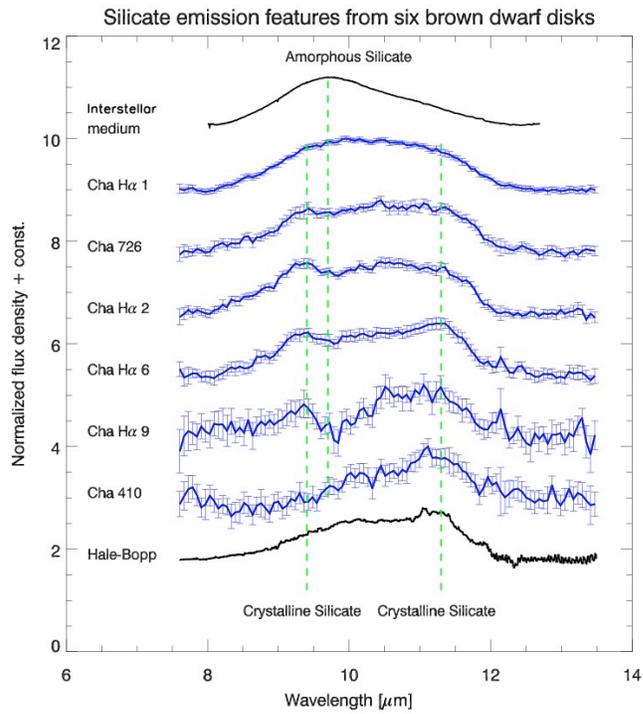

**Fig. 1** Continuum-subtracted and normalized silicate emission features from our targets. For comparison the spectra of the amorphous silicate-dominated interstellar medium and the crystalline-rich comet Hale-Bopp (*5*) are also shown. The 9.3 μm peak is mainly enstatite, the 11.3 μm peak is from forsterite.

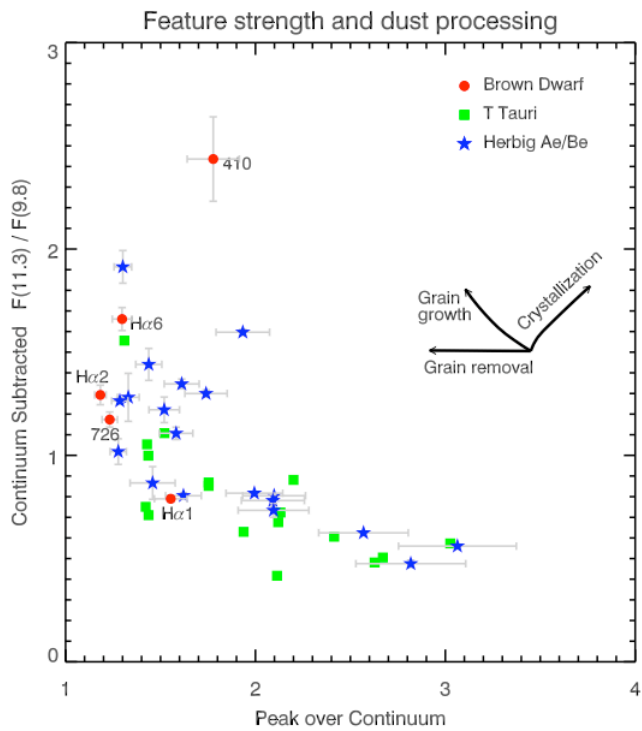

**Fig. 2** Crystalline contribution to the silicate feature (flux at 11.3 μm over flux at 9.8 μm) as a function of the emission feature strength (peak flux over continuum flux). The correlation recognized for intermediate- and low-mass young stars (Herbig Ae and T Tauri) holds for brown dwarfs, but it is not linear.

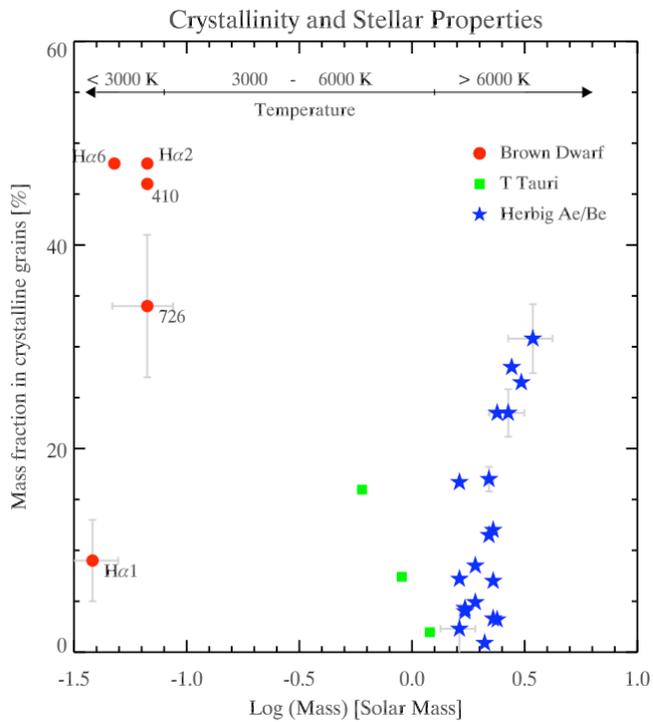

**Fig. 3**  Stellar and substellar mass and temperature as a function of the crystalline mass fraction. Measurements of the brown dwarf disks probe the relation over a significantly broader parameter range than previous observations. The suggestive decline of crystallinity with stellar mass apparent for Herbig Ae/Be stars is not a general trend: even very low mass and cool objects can harbor highly crystallized disks. Typical uncertainties are indicated.

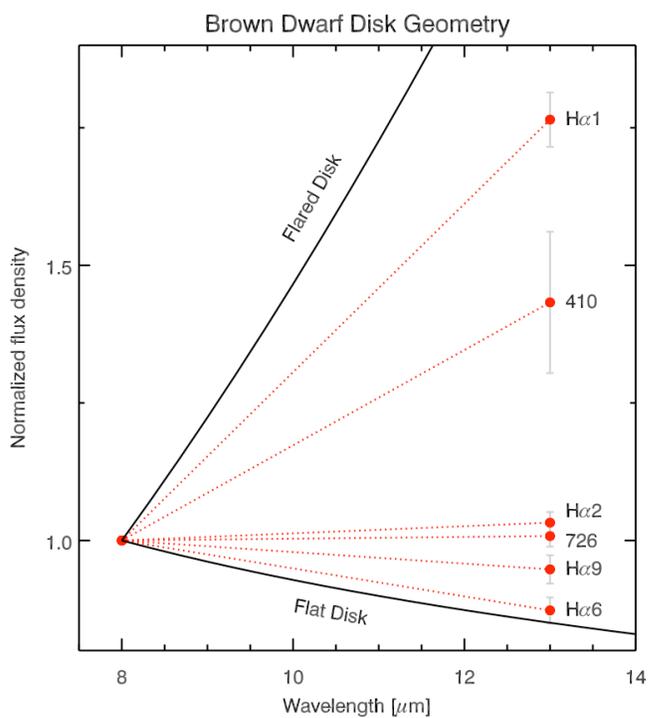

**Fig. 4** Dust settling in six brown dwarf disks. The flux densities of the disks at 8 and 13 μm are normalized to the 8 μm flux and plotted against the wavelength. The slopes of the spectral energy distributions for flat and flared disks are overplotted. All disks show intermediate flaring, four disks are close to the flat geometry.